# Surgical task expertise detected by a self-organizing neural network map

B. Dresp-Langley[1*], R. Liu[2], and J. M. Wandeto[3]

[1] *Centre National de la Recherche Scientifique ICube UMR 7357, Strasbourg University, Strasbourg, France*
[2] *Robotics Department ICube UMR 7357, Strasbourg University, Strasbourg, France*
[3] *Department of Information Technology, Dedan Kimathi University of Technology, Nyeri, Kenya*
[*] *Correspondence, email: birgitta.dresp@unistra.fr*

*Abstract: Individual grip force profiling of bimanual simulator task performance of experts and novices using a robotic control device designed for endoscopic surgery permits defining benchmark criteria that tell true expert task skills from the skills of novices or trainee surgeons. Here we show that grip variability in a true expert and a complete novice executing a robot-assisted surgical simulator task reveal statistically significant differences as a function of task expertise, predicted by the output metric of a Self-Organizing neural network Map (SOM) with a bio-inspired functional architecture that maps the functional connectivity of the somatosensory neural networks of the primate brain.*

## I. Introduction

Current state of the art in robotic assistance for surgical procedures [1,2] has a considerable potential for augmenting the precision and capability of physicians, but technological challenges still need to be met in terms of optimized system architecture, software, mechanical design, imaging systems, and user interface design and management for maximum safety. Moreover, objective quantitative performance criteria need to be worked out for defining gold standards of true expert performance in this emerging realm of assistive technology for pushing optimal training programs for novice surgeons [3,4]. In Previous work by ourselves and other [3-8] has exploited sensor data and, most recently, wireless wearable sensor technology to demonstrate how individual grip force profiling of bimanual simulator task performance of experts and novices using a robotic control device designed for endoscopic surgery permits to find benchmark criteria for telling true expert task skills from the skills of novices or trainees. Important universal criteria for expert performance are a stable speed-precision trade-off aimed at maximal precision, not speed, while executing a surgical task [9], and minimized variability in performance scores relative to precision [8-10], task execution speed [3-5,8-10], and hand grip forces [4-7], which by definition exhibit optimal prehensile synergy in a true expert. In this work here we show that 1) the variability of the bimanual grip forces of a true expert and a complete novice executing a robot-assisted surgical simulator task reveals a statistically significant difference as a function of task expertise, and 2) this difference is captured by a SOM with a bio-inspired functional architecture that maps the functional connectivity of the somatosensory hand-to-brain-and-back circuitry in the human primate [11]. The data-driven approach has potential for a parsimonious, economic, and functionally meaningful automated analysis of surgical task skill evolution.

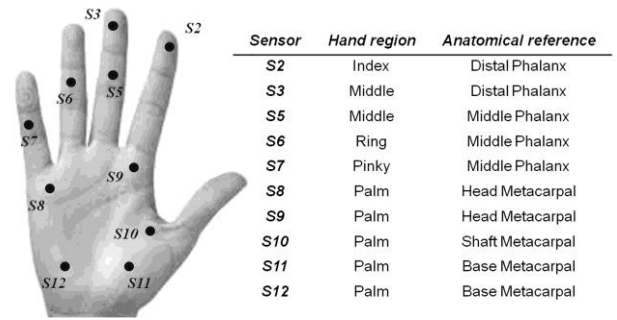

*Figure 1: Sensor locations corresponding to mechanoreceptor regions generating thousands of grip forces ( expert and novice data), exploited and modeled here.*

## II. Material and methods

The **robotic task system**, the simulator task, and the wearable wireless grip force sensor gloves used here are described in full detail in [4,5]. We analyzed a total of 239 710 grip force data sampled at the millivolt (mV) scale every 20 milliseconds from a surgical expert and a complete novice performing the robot-assisted simulator task in ten repeated sessions with their dominant or non-dominant hands. The **neural network architecture** exploited for modeling the grip force data here follows from some of our previous work [12-14] on functional properties of the Quantization Error (QE) in the output of a Self-Organizing Map (SOM), which is described formally as a nonlinear, ordered, smooth mapping of high-dimensional input data onto the elements of a regular, low-dimensional array. The set of input variables is definable as a real vector $x$, of n-dimension. With each element in the SOM array we associate a parametric real vector $m_i$ of n-dimension as a model. Assuming a general distance measure between $x$ and $m_i$ denoted by d($x$, $m_i$), the map of an input vector $x$ on the SOM array is defined as the array element $m_c$ that best matches $x$ (smallest d($x$, $m_i$)). During the learning process, models topographically close in the map up to a certain geometric distance, denoted by $h_{ci}$ will activate each other to learn something from their shared input $x$. This will result in a local relaxation or smoothing effect on the models in this neighborhood, which in continued learning leads to global ordering. SOM learning is represented by the equation

$$m(t+1) = m_i(t) + \alpha(t)\, h_{ci}(t)\, [x(t) - m_i(t)] \quad (1)$$

where t =1,2,3...is an integer, the discrete-time coordinate, $h_{ci}(t)$ is the neighborhood function, a smoothing kernel defined over the map points which converges towards zero with time, $\alpha(t)$ is the learning rate. At the end of the *winner-take-all* learning process, each input vector $x$ becomes associated to its best matching model on the map mc. The difference between $x$ and $m_c$, $\|x - m_c\|$ is reflected by the quantization error QE. The QE of $x$ is given by

$$QE = 1/N \sum_{i=1}^{N} \|x_i - m_{c_i}\| \quad (2)$$

where $N$ is the number of input vectors $x$. The SOM implemented to map the mechanoreceptor-to-brain network for this study here was a 7 by 7 map creating a fully connected network of 49 neurons where each of the ten sensors from which data were exploited here (Fig. 1) contributes to the synaptic weight of each neuron. The QE in the SOM output (SOM-QE) is used to model the variability in thousands of grip force data of a true expert and a novice.

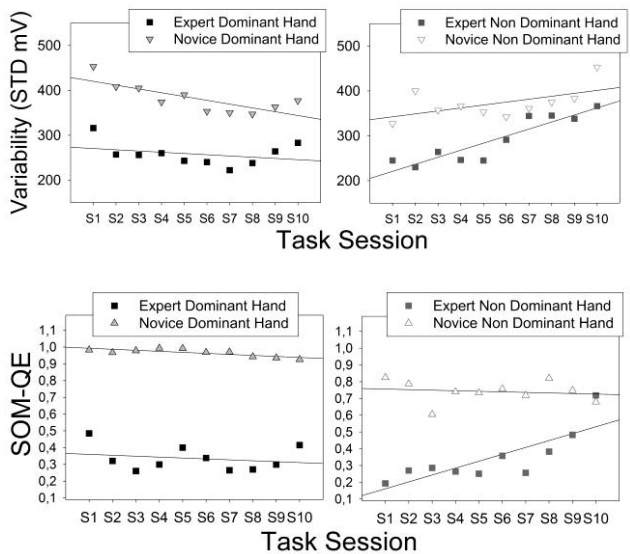

*Figure 2: The variability (STD in mV) of individual grip forces in the task sessions of a true expert and a complete novice is reliably predicted by a functionally pertinent neural network metric from the brain-inspired Self-Organizing Map (SOM-QE).*

## III. Results and discussion

Data variability was computed in terms of the standard deviations (STD) of the means per condition and session (Figure 2, top). The SOM-QE from the neural network analyses of the same data for each condition and session was computed (Figure 2, bottom). Further statistical analyses of the data yield significant effects of task expertise on STD (t(1,18)=22.34; p<.001 for dominant; t(1,18)=7.43; p<001 for non-dominant), mirrored by similar significant effects on the SOM-QE (t(1,18)=9.27; p<.001 for dominant; t(1,18)=4.09; p<.001 for non-dominant), showing that task skill-related grip forces are reliably predicted by the brain-inspired SOM-QE.

## IV. Conclusion

Combining grip force sensor technology with predictive modeling [13-15] by Artificial Intelligence, as shown here, promises for an economic, functionally meaningful automated analysis of surgical task skill evolution.

**ACKNOWLEDGMENTS**
This work is part of the IDEX (Initiative D'EXcellence) of Strasbourg University. Support of the CNRS is gratefully acknowledged.

**AUTHOR'S STATEMENT**
Authors state no conflict of interest. Informed consent has been obtained from all individuals included in this study. The research complies with all relevant national regulations or institutional policies, performed in accordance with the tenets of the Helsinki Declaration, and has been approved by the authors' institutional review board or equivalent.